\documentclass[12pt]{revtex4}
\usepackage{dcolumn}
\usepackage{graphicx}
\usepackage{amsmath}
\usepackage{amsfonts}
\usepackage{amssymb}
\usepackage{psfrag}
\usepackage{wrapfig}
\usepackage{subfigure}
\usepackage{makeidx}
\usepackage{bm}
\usepackage{epsf}
\usepackage{hyperref}

\begin{document}

\title{Spherically symmetric gravitational collapse of a dust cloud
in third order Lovelock Gravity}

\author{ Kang Zhou$^{1}$, Zhan-Ying Yang$^{1}$\footnote{Email:zyyang@nwu.edu.cn},
De-Cheng Zou$^{1}$ and Rui-Hong Yue$^{2}$\footnote{ Email:yueruihong@nbu.edu.cn}}

\affiliation{ $^{1}$Department of Physics, Northwest University, Xi'an, 710069, China\\
$^{2}$Faculty of Science, Ningbo University, Ningbo 315211, China}
\date{\today}

\begin{abstract}
\indent

We investigate the spherically symmetric gravitational collapse of
an incoherent dust cloud by considering a LTB type
space-time in third order Lovelock Gravity without cosmological constant,
and give three families of LTB-like solutions which separately
corresponding to hyperbolic, parabolic and elliptic.
Notice that the contribution
of high order curvature corrections have a profound influence on the
nature of the singularity, and the global structure of space-time
changes drastically from the analogous general relativistic case.
Interestingly, the presence of high order Lovelock terms leads to
the formation of massive, naked and time-like singularities
in the 7D space-time, which
is disallowed in general relativity. Moveover, we point out that the naked
singularities in the 7D case may be gravitational weak therefore may not be a
serious threat to the cosmic censorship hypothesis, while the naked
singularities in the $D\geq8$ inhomogeneous collapse violate the cosmic
censorship hypothesis seriously.
\end{abstract}



\maketitle

\section{Introduction\label{intro}}
\indent

One of the most intriguing predictions of Einstein's general
relativity is the existence of black holes which are formed from
gravitational collapse in the last stage of heavy stars' life or in
high-density regions of the density perturbations in the early
universe \cite{Oppenheimer:1939ue}. The gravitational
collapse of an incoherent spherical dust cloud is
described by the  Einstein equation
$G_{ab}=\varepsilon(t,r)u_au_b$  with  the energy density
$\varepsilon(t,r)$  and the time-like velocity vector  $u_a$.
In  a reference frame comoving with the collapsing matter,
it can be proved that the metric of spherically symmetric space-time
reads \cite{Lemaitre:1933gd}
\begin{eqnarray}
ds^2=-dt^2+A(t,r)^2dr^2+R(t,r)^2(d\theta^2+\sin^2\theta d\phi^2),\nonumber\\\label{eq:1a}
\end{eqnarray}
where $r$ is the comoving radial coordinate and $t$  the proper time
of freely falling shells. As well known, the solution of this metric
is the Lemaitre-Tolman-Bondi (LTB) solution \cite{Lemaitre:1933gd},
which has been extensively studied not only in spherical collapse, but
also in cosmology as well \cite{Maartens:1995dx}\cite{Garfinkle:2006sb}.

It is well known that the end state of gravitational collapse is
a singularity of space-time
curvature with infinite density of matter\cite{Hawking:1969vz}\cite{Hawking:1973uf}.
But, it is not known
clearly whether such a singularity will be naked or covered by a
event horizon. At the singularity, the laws of science and the
ability to predict the future would break down. It led Roger
Penrose to propose the (weak and strong) cosmic censorship
hypothesis (CCH) \cite{Hawking:1973uf}\cite{Penrose:1969}, which to date
remain unproven. The weak CCH asserts that there can be no
singularity visible from future null infinity and light from
singularity is completely blocked by the event horizon. It protects
external observers from the consequences of the breakdown of
predictability that occurs at the singularity, but dose nothing at
all to observers who are cofalling with the collapsing massive body,
while the strong CCH prohibits singularity's visibility by any
observer. Despite almost 40 years of effort we still don't have a
general proof of CCH \cite{Joshi:2000fk}. On the contrary,
it was shown that the seminal work of Oppenheimer and Snyder
\cite{Oppenheimer:1939ue} is not a typical model and the
central singularities formed in generic collapse are naked and observable
\cite{Christodoulou:1984mz}\cite{Nolan:2002mz}.

Up to now, many quantum theories to unify all fundamental interactions
including gravity have been proposed, the
most promising candidate among them is   superstring/M-theory. In string
theory, extra dimensions were promoted from an interesting curiosity
to a theoretical necessity since superstring theory requires a
ten-dimensional space-time to be consistent from the quantum
point of view \cite{Horava:1996ma}\cite{Randall:1999ee}. It caused a
renewed interest about the general relativity in more than 4
dimensions. Several LTB-like solutions to Einstein equations in
higher dimensions have been obtained in recent years
\cite{Banerjee:1994us}\cite{Ghosh:2001fb}.
On the other hand, in recent
years another renewed interest has grown in higher order gravity,
which involves higher derivative curvature terms. Among the higher curvature
gravities, the most extensively studied theory is so-called Lovelock
gravity \cite{Lovelock:1971yv}, which   naturally emerged when we
want to generalize Einstein's theory in higher dimension by
keeping  all characteristics of usual general relativity
excepting the linear dependence of Riemann tensor. In addition,  Lovelock terms naturally occur in the effective low-energy action of
superstring theory \cite{Lust:1989tj}\cite{Elizalde:2007pi}. Hence, the Lovelock gravity provides a promising
framework to study curvature corrections to the Einstein-Hilbert action. Since we are interested in what influence
to the theory of gravity will be caused by Lovelock terms, for simplicity,
we only consider pure Lovelock gravity and neglect the other terms in the
low-energy expansion string theory.
The Lagrangian of Lovelock theory is the sum of dimensionally
extended Euler densities \cite{Lovelock:1971yv}
\begin{eqnarray}
{\cal L}=\sum^{m}_{n=0}\alpha_n{\cal L}_n,\nonumber
\end{eqnarray}
where arbitrary constants $\alpha_n$ are the Lovelock coefficients,
and ${\cal L}_n$ is the Euler density of a $2k$-dimensional manifold
\begin{eqnarray}
{\cal L}_n=\frac{1}{2^n}\delta^{a_1b_1\cdots a_nb_n}_{c_1b_1\cdots
c_nd_n} R^{c_1d_1}_{~~~~a_1b_1}\cdots
R^{c_nd_n}_{~~~~a_nb_n}.\label{eq:2a}
\end{eqnarray}
Here the generalized delta function $\delta^{a_1\cdots
b_n}_{c_1\cdots d_n}$ is totally antisymmetric in both sets of
indices and $R^{cd}_{~~ab}$ is the Riemann tensor. Though the
Lagrangian of Lovelock gravity consists of some higher derivative
curvature terms, its field equations of motion contain the most
symmetric conserved tensor with no more than two derivative of the
metric. The ${\cal L}_0 $ is assumed to be identity and  $c_0$  the cosmological
constant. ${\cal L}_1$ gives the usual Einstein-Hilbert action term.
${\cal L}_2$ is called Gauss-Bonnet term, it is a correction term of
the action \cite{Bento:1995qc}. So far, the exact static and
spherically symmetric black hole solutions in third order Lovelock
gravity were first found in \cite{Myers:1989kt},
and the thermodynamics have been investigated
in \cite{Ge:2009ac}\cite{Dehghani:2009zzb}\cite{Dehghani:2005zzb}.

It should be interesting to discuss the gravitational collapse in
higher dimensional Lovelock gravity. The natural questions would be, for
instance, how does the Lovelock terms affect final fate of collapse?
What horizon structure will be formed? Whether solutions leads to
naked singularities? Whether CCH can be hold?
Recently, Maeda considered the
spherically symmetric gravitational collapse of a inhomogeneous dust
with the $D \ge 5$-dimensional action including the Gauss- Bonnet
term. He discussed its effects on the final fate of gravitational
collapse without finding the explicit form of the solution \cite{Maeda:2006pm}.
Then, Jhingan and Ghosh considered the 5D
action with the Gauss-Bonnet terms for gravity and give a exact
model of the gravitational collapse of a inhomogeneous dust
\cite{Jhingan:2010zz}\cite{Ghosh:2010jm}. It's interesting to investigate
the effect of higher order terms of Lovelock gravity in gravitational collapse.
we explore the gravitational collapse in comoving coordinates, and
seek LTB-like solutions in the third order Lovelock gravity. The dimensions
of space-time concerned is $D\geq7$ because the third order
term yields nontrivial effect in dimensions greater than or
equal to 7 \cite{Dehghani:2006yd}\cite{Dehghani:2005vh}.
Using our solutions, we discuss the
formation of singularities, and analyze the nature of them.
In particular, we consider that whether such singularities
would be hidden or be visible to outside observers. Since
the CCH would be violate by naked gravitational strong singularities,
we investigate the gravitational strength of naked singularities,
and discuss that whether the naked singularities in third order
theory is a serious threat to CCH.

This paper is organized as follows. In Sec \ref{22s}, for
the $D\geq7$-dimensional space-time, we give the field equations in third order
Lovelock gravity without a cosmological constant, and derive the
LTB-like solutions. In Sec \ref{33s}, we investigate the final fate of
the spherically symmetric gravitational collapse of a dust cloud.
The subject of Sec \ref{44s} is to analyze horizons (both apparent horizons and event horizons) and
trapped surfaces, and explore that whether the singularity formed by
gravitational collapse is hidden or visible to outside observers.
The strength of the singularity is demonstrated in Sec \ref{55s}. Sec
\ref{66s} is devoted to conclusions and discussions.

Throughout this paper we use units such that $8\pi G = {c^4} = 1$.

\section{ LTB-like solutions in third order
Lovelock gravity \label{22s}}
\indent

The action of third order Lovelock gravity in the presence of matter field can
be written as
\begin{eqnarray}
S=\int{d^D x\sqrt{-g}(R+\alpha_2{\cal L}_2+\alpha_3{\cal L}_3)}+S_{matter},\label{eq:3a}
\end{eqnarray}
where $\alpha _2$ and $\alpha_3$ are coupling constants of the
second order (Gauss-Bonnet) and the third order terms, respectively. In the low-energy limit of the heterotic string theory,
$\alpha$ is regard as the inverse string tension and positive define \cite{Boulware:1985wk}. Hence we restrict ourselves to the case $\alpha\geq0$ in this paper.
For future simplicity, we take coefficients $\alpha _2=\frac{\alpha
}{(D-3)(D-4)}$ and $\alpha _3=\frac{\beta}{72C_{D-3}^4}$.
The Gauss-Bonnet term ${\cal L}_2$ is
\begin{eqnarray}
{\cal L}_2 &=&R_{\mu\nu\sigma\kappa}R^{\mu\nu\sigma\kappa}-4R_{\mu\nu}R^{\mu\nu}+R^2\nonumber
\end{eqnarray}
and the third order terms of Lovelock Lagrangian is of the form
\begin{eqnarray}
{\cal L}_3&=&2R^{\mu\nu\sigma\kappa}R_{\sigma\kappa\rho\tau}R^{\rho\tau}_{~~\mu\nu}
+8R^{\mu\nu}_{~~\sigma\rho}R^{\sigma\kappa}_{~~\nu\tau}R^{\rho\tau}_{~~\mu\kappa}\nonumber\\
&+&24R^{\mu\nu\sigma\kappa}R_{\sigma\kappa\nu\rho}R^{\rho}_{\mu}+
3RR^{\mu\nu\sigma\kappa}R_{\mu\nu\sigma\kappa}\nonumber\\
&+&24R^{\mu\nu\sigma\kappa}R_{\sigma\mu}R_{\kappa\nu}
+16R^{\mu\nu}R_{\nu\sigma}R^{\sigma}_{~\mu}\nonumber\\
&-&12RR^{\mu\nu}R_{\mu\nu}+R^3.\nonumber
\end{eqnarray}
Varying the action Eq.~(\ref{eq:3a}), we obtain the equation of gravitation field
\begin{eqnarray}
G_{\mu\nu}^{(1)} + \alpha _2G_{\mu\nu}^{(2)}+\alpha_3G_{\mu\nu}^{(3)}=T_{\mu\nu},\label{eq:4a}
\end{eqnarray}
where
\begin{eqnarray}
G_{\mu\nu}^{(1)}&=&{R_{\mu\nu}} - \frac{1}{2}R{g_{\mu\nu}}, \nonumber\\
G_{\mu\nu}^{(2)}&=&2(R_{\mu\sigma\kappa\tau}R_{\nu}^{~\sigma\kappa\tau}
-2R_{\mu\rho\nu\sigma}R^{\rho\sigma}-2R_{\mu\sigma}R^{\sigma}_{~\nu}+RR_{\mu\nu})\nonumber\\
& -& \frac{1}{2}{\cal L}_2{g_{\mu\nu}},\nonumber\\
G^{(3)}_{\mu\nu}&=&3R_{\mu\nu}R^2-12RR_{\mu}^{~\sigma}R_{\sigma\nu}
-12R_{\mu\nu}R_{\alpha\beta}R^{\alpha\beta}\nonumber\\
&+&24R_{\mu}^{~\alpha}R_{\alpha}^{~\beta}R_{\beta\nu}-24R_{\mu}^{~\alpha}R^{\beta\sigma}R_{\alpha
\beta\sigma\nu}\nonumber\\
&+&3R_{\mu\nu}R_{\alpha\beta\sigma\kappa}R^{\alpha\beta\sigma\kappa}
-12R_{\mu\alpha}R_{\nu\beta\sigma\kappa}R^{\alpha\beta\sigma\kappa}\nonumber\\
&-&12RR_{\mu\sigma\nu\kappa}R^{\sigma\kappa}
+6RR_{\mu\alpha\beta\sigma}R_{\nu}^{~\alpha\beta\sigma}\nonumber\\
&+&24R_{\mu\alpha\nu\beta}R_{\sigma}^{~\alpha}R^{\sigma\beta}+24R_{\mu\alpha\beta\sigma}R_{
\nu}^{~\beta}R^{\alpha\sigma}\nonumber\\
&+&24R_{\mu\alpha\nu\beta}R_{\sigma\kappa}R^{\alpha\sigma\beta\kappa}
-12R_{\mu\alpha\beta\sigma}R^{\kappa\alpha\beta\sigma}R_{\kappa\nu}\nonumber\\
&-&12R_{\mu\alpha\beta\sigma}R^{\alpha\kappa}R_{\nu\kappa}^{~~\beta\sigma}
+24R_{\mu}^{~\alpha\beta\sigma}R_{\beta}^{~\kappa}R_{\sigma\kappa\nu\alpha}\nonumber\\
&-&12R_{\mu\alpha\nu\beta}R^{\alpha}_{~\sigma\kappa\rho}R^{\beta\sigma\kappa\rho}
-6R_{\mu}^{~\alpha\beta\sigma}R_{\beta\sigma}^{~~\kappa\rho}R_{\kappa\rho\alpha\nu}\nonumber\\
&-&24R_{\mu\alpha}^{~~\beta\sigma}R_{\beta\rho\nu\lambda}R_{\sigma}^{~\lambda\alpha\rho}
-\frac{1}{2}{\cal L}_3g_{\mu\nu}.\nonumber
\end{eqnarray}

The solution we search is collapse of a spherically symmetric dust
in $D\geq7$ space-time. Following LTB solutions, we assume that the system
consists of a freely falling perfect fluid, it requires that the mean
free path between collisions is small compared with the scale of
lengths used by observer. The energy-momentum tensor for
the perfect fluid in comoving coordinates is
\begin{eqnarray}
{T_{\mu\nu}} = \varepsilon ( {t,r} )u_\mu u_\nu,\label{eq:5a}
\end{eqnarray}
where ${u_\mu} = \delta _\mu^t$ is the velocity vector field. We
assume that the energy density $\varepsilon(t,r)$ on the initial
surface is smooth, that is, it is can be extended to a $C^\infty$ function
 on the entire real line. The metric in comoving coordinates is written
in the form
\begin{eqnarray}
ds^2 =  - dt^2 + A( {t,r} )^2dr^2 + R( {t,r} )^2d\Omega
_{D-2}^2,\label{eq:6a}
\end{eqnarray}
where  $r$ is the comoving radial cooedinate, and $t$ is
the proper time of freely falling shells. Plugging the metric Eq.~(\ref{eq:6a}) into Eq.~(\ref{eq:4a}), we have
\begin{eqnarray}
G_r^t &=& \frac{2-D}{{A^5}{R^5}}(\dot{A}R' - A\dot{R}')[\beta(A^2\dot{R}^2+A^2-R'^2)^2\nonumber\\
&+&2\alpha A^2R^2(A^2\dot{R}^2+A^2-R'^2)+A^4R^4]= 0, \label{eq:7a}
\end{eqnarray}
where an over-dot and prime denote the partial derivative with
respect to $t$ and $r$, respectively. Based on Eq.~(\ref{eq:7a}),
we arrive at two families of solutions which satisfy
\begin{eqnarray}
0=\dot{A}R' - A\dot{R}',\label{eq:8aa}
\end{eqnarray}
and
\begin{eqnarray}
0&=&\beta(A^2\dot{R}^2+A^2-R'^2)^2\nonumber\\
&+&2\alpha A^2R^2(A^2\dot{R}^2+A^2-R'^2)+A^4R^4,\label{eq:9a}
\end{eqnarray}
respectively. The second equation involves the Lovelock coupling constants,
and  leads  to a trivial solution
if  $\alpha \to 0, \beta\to 0$. We will neglect it
since we want to explore the Lovelock corrections to Einstein gravity. The solution of Eq.(\ref{eq:8aa}) reads
\begin{eqnarray}
A(t,r) = \frac{R'(t,r)}{W(r)}\label{eq:8a}
\end{eqnarray}
with an arbitrary function $W( r )$.

As it is possible to make an arbitrary re-labeling of
spherical dust shell by $r \to g( r )$, we fix the labeling by
requiring that, on the hypersurface $t = 0$, $r$ coincides with the
area radius
\begin{eqnarray}
R( {0,r} ) = r.\label{eq:10a}
\end{eqnarray}
Apparently, every $t=const$ and $r = const$ slice of the
space-time is a sphere of radius $R$. The radius $R$ can be given
an absolute significance by following interpretation. Considering two
particles $a$ and $b$ distribution along the radial direction in
comoving coordinates, the space distance between them at the
coordinate time $t$ can be obtained on the hypersurface $t=const$ as
\begin{eqnarray}
{l_{ab}} &=& \int_a^b {dl} =\int_a^b \sqrt{{h_{ij}}dx^i dx^j} \nonumber\\
&=&\int_{r_a}^{r_b}\frac{R'}{\sqrt{1+K}}dr=\int_{{R_a}}^{{R_b}}
{\frac{{d{R}}}{\sqrt{1+K}}},\label{eq:11a}
\end{eqnarray}
here ${h_{ab}}$ is the induced metric on the hypersurface, $K( r ) = W{( r )^2} - 1$ and we assume $K(r) > -1$. In last step the fact $
d{R} = R'dr + \dot{R}dt = R'dr$ at a constant $t$  is used.
Thus, the line element of actual space distance along radial
direction is ${\frac{{dR}}{\sqrt{1+K}}}$. The signature of $K( r )$
corresponds  to three types of
solutions, namely hyperbolic, parabolic and
elliptic, respectively. $K( r ) = 0$ is the marginally bound case in
which the metric takes the form of Minkowski metric
on the hypersurface $t = 0$.

In comoving coordinates, the equations of momentum conservation $(
T_i^\mu)_{;\mu }=0$ are automatically satisfied, and the $t$-component reads
\begin{eqnarray}
0&=&-\frac{\partial \varepsilon}{\partial t}-\varepsilon(
\frac{(R'^2)_{,t}}{2R'^2}+\frac{(D-2)(R^2)_{,t}}{2R^2})\nonumber\\
&=&-\frac{\partial }{{\partial t}}( {\varepsilon {R^{D-2}}R'} ),\label{eq:12a}
\end{eqnarray}
which gives the solution
\begin{eqnarray}
\varepsilon (t,r)&=&\frac{\varepsilon (0,r)r^{D-2}}{R^{D-2}R'}. \label{eq:13a}
\end{eqnarray}
Thus, the mass function is defined as
\begin{eqnarray}
M(r)&=&\frac{6}{D - 2}\int \varepsilon(t,r)R^{D - 2}d{R}\nonumber\\
&=&\frac{6}{D- 2}\int_0^r \varepsilon(0,r)r^{D - 2}dr,\label{eq:14a}
\end{eqnarray}
it is  positive and increases with increasing $r$ for the
nonnegativity of energy density $\varepsilon $.

Based on Eq.~(\ref{eq:8a}) and $K( r ) = W{( r )^2} - 1$,
the $tt$ component of equations is given by
\begin{eqnarray}
G_t^t &=& \frac{2-D}{6R^{D-2}R'}\Big[R^{D-7}\Big(\beta(\dot{R}^2-K)^3
+ 3\alpha (\dot{R}^2-K)^2{R^2}\nonumber\\
&+&3(\dot{R}^2-K){R^4} \Big)\Big]'
=- \varepsilon (t,r).\label{eq:15a}
\end{eqnarray}
Substituting Eq.~(\ref{eq:13a}) and Eq.~(\ref{eq:14a}) into Eq.~(\ref{eq:15a}), we get the real solution that
\begin{eqnarray}
{\dot{R}^2} = K+\frac{R^2}{\beta}[(\frac{\Theta+\Pi}{2})^{\frac{1}{3}}+(\frac{\Theta-\Pi}{2})^{\frac{1}{3}}-\alpha],\label{eq:001c}
\end{eqnarray}
where
\begin{eqnarray}
\Theta&=&-2\alpha^3+3\alpha\beta +\beta^2\rho_D,\nonumber\\
\Pi&=&\sqrt{\beta^2(\beta^2 \rho_D^2-4\alpha^3\rho_D+6\alpha\beta \rho_D-3\alpha^2+4\beta) }\nonumber
\end{eqnarray}
with $\rho_D=MR^{1-D}$. This equation governs the time evolution of $R$ in D-dimensional third Lovelock gravity.
For gravitational collapse, the solution of  $R(t,r)$ takes
\begin{eqnarray}
\dot{R}=  - \sqrt { K+\frac{R^2}{\beta}\Big[\Big(\frac{\Theta+\Pi}{2}\Big)^{\frac{1}{3}}+\Big(\frac{\Theta-\Pi}{2}\Big)^{\frac{1}{3}}-\alpha\Big]}.\label{eq:02c}
\end{eqnarray}
It is straightforward to check that other field equations are automatically
satisfied when Eq.~(\ref{eq:001c}) is satisfied.

For arbitrary initial data of energy density $\varepsilon(0,r)$,
the Eq.~(\ref{eq:02c}) completely specify the dynamical evolution of
collapsing dust shells. In the general relativistic limit $\alpha\to 0,\beta\to 0$
with marginally bound case, Eq.~(\ref{eq:02c}) can be
integrated to yield
\begin{eqnarray}
{t_b}( r ) - t( r ) = \frac{2\sqrt {3}R^{\frac{D-1}{2}}}{(D-1)\sqrt {M( r )}},\label{eq:18a}
\end{eqnarray}
where $t_b(r)$ is an function of integration, and can be formulated as
\begin{eqnarray}
{t_b}( r ) = \frac{2\sqrt {3}r^{\frac{D-1}{2}}}{(D-1)\sqrt{M( r )}}.\nonumber
\end{eqnarray}
Consequently, the function $R( {t,r} )$ is
\begin{eqnarray}
R(t,r) = r{\Big[ 1 + \frac{D-1}{2\sqrt {3}r^{\frac{D-1}{2}}}t\sqrt{M(r)}
\Big]^{\frac{2}{D-1}}}.\label{eq:20a}
\end{eqnarray}
It is similar to the form of $R( {t,r} )$ in LTB and
LTB-like-solutions \cite{Lemaitre:1933gd}\cite{Ghosh:2001fb}.

We can consider that our LTB-like solution is attached at the boundary of the dust cloud, which is represented by a finite constant comoving radius $r=r_0>0$, to the outside vacuum region. The outside vacuum region is represented by the solution whose metric is
\begin{eqnarray}
ds^2=-F(\tilde{r})dT^2+\frac{d\tilde{r}^2}{F(\tilde{r})}+\tilde{r}^2d\Omega_{D-2}^2,\nonumber
\end{eqnarray}
where $F(\tilde{r})$ takes form of
\begin{eqnarray}
F(\tilde{r})=1-\frac{\tilde{r}^2}{\beta}\Big[\Big(\frac{\tilde{\Theta}+\tilde{\Pi}}{2}\Big)^{\frac{1}{3}}
+\Big(\frac{\tilde{\Theta}-\tilde{\Pi}}{2}\Big)^{\frac{1}{3}}-\alpha\Big],\label{eq:03c}
\end{eqnarray}
where
\begin{eqnarray}
\tilde{\Theta}&=&-2\alpha^3+3\alpha\beta +\beta^2\tilde{\rho}_D,\nonumber\\
\tilde{\Pi}&=&\sqrt{\beta^2(\beta^2 \tilde{\rho}_D^2-4\alpha^3\tilde{\rho}_D+6\alpha\beta \tilde{\rho}_D-3\alpha^2+4\beta) }\nonumber
\end{eqnarray}
with $\tilde{\rho}_D=\frac{6m}{(D-2)\Omega_{D-2}}\tilde{r}^{1-D}$.
If we define that
\begin{eqnarray}
m&=&\frac{(D-2)\Omega_{D-2}M}{6},\nonumber\\
R&=&\tilde{r},\nonumber\\
dT&=&\frac{\sqrt{1+K(r)}}{F(R)}dt,\label{eq:04c}
\end{eqnarray}
we can prove that the LTB-like solution attached at a finite constant comoving radius $r=r_0$, where we represent this hypersurface as $\Sigma$, to the outside vacuum solution smoothly.

{\it Proof:} As seen from inside of $\Sigma$, the metric on $\Sigma$ is obtained by
\begin{eqnarray}
ds_{\Sigma}^2=-dt^2+R_{\Sigma}^2d\Omega_{D-2}^2.\nonumber
\end{eqnarray}
As seen from outside of $\Sigma$, it is
\begin{eqnarray}
ds_{\Sigma}^2=-(F(R_{\Sigma})\dot{T}_{\Sigma}^2-\frac{\dot{R}_{\Sigma}^2}{F(R_{\Sigma})})dt^2+R_{\Sigma}^2d\Omega_{D-2}^2.\nonumber
\end{eqnarray}
It can be checked that the induced metric is the same on both sides of the hypersurface $\Sigma$ with the definition Eq.~(\ref{eq:04c}). It implies
\begin{eqnarray}
F(R_{\Sigma})\dot{T}_{\Sigma}^2-\frac{\dot{R}_{\Sigma}^2}{F(R_{\Sigma})}=1.\label{eq:05c}
\end{eqnarray}
We can define a function $\zeta$ by
\begin{eqnarray}
\zeta\equiv\sqrt{F(R_{\Sigma})+\dot{R}_{\Sigma}^2}=F(R_{\Sigma})\dot{T}_{\Sigma}.\nonumber
\end{eqnarray}
As seen from inside, the nonzero components of the extrinsic curvature ${\cal K}_a^b$ of $\Sigma$ are calculated as ${\cal K}_t^t=0$ and ${\cal K}_i^i=\frac{\sqrt{1+K(r_0)}}{R_{\Sigma}}$. As seen from outside of $\Sigma$, we find ${\cal K}_t^t=\frac{\dot{\zeta}}{\dot{R}_{\Sigma}}(F(R_{\Sigma})\dot{T}_{\Sigma}^2-\frac{\dot{R}_{\Sigma}^2}{F(R_{\Sigma})})^{-1}$ and ${\cal K}_i^i=\frac{\zeta}{R_{\Sigma}}$. With the definition Eq.~(\ref{eq:04c}), ${\cal K}_a^b$ is the same on both sides of $\Sigma$. It implies
\begin{eqnarray}
\zeta=\sqrt{1+K(r_0)}.\label{eq:06c}
\end{eqnarray}
Finally, we combine Eq.~(\ref{eq:05c}) and Eq.~(\ref{eq:06c}) giving the equation of motion for the hypersurface $\Sigma$ as
\begin{eqnarray}
\dot{R}_{\Sigma}^2=1-F(R_{\Sigma})+K(r_0).\nonumber
\end{eqnarray}
It takes the same form as Eq.~(\ref{eq:01c}) with the definition Eq.~(\ref{eq:04c}). Thus, the LTB-like solution attached at a finite constant comoving radius $r=r_0$ to the outside vacuum solution smoothly. $\Box$

\section{Shell focusing singularity\label{33s}}
\indent

In this section, we consider the final fate of the gravitational
collapse in $D\geq7$ space-time. As pointed earlier, in general
relativity, a collapse leads to a singularity, and the conjecture
that such a singularity must be covered by an event horizon is the
weak CCH. There are two kinds of singularities: shell
crossing singularity and shell focusing singularity which is defined by
$R' = 0$ and $R = 0$, respectively. The characteristic of a
singularity in space-time manifold is the divergence of the Riemann
tensor and the energy density \cite{Hawking:1973uf}. In our case,
the Kretschmann invariant
scalar ${\cal K} = {R_{\mu\nu\sigma\tau}}{R^{\mu\nu\sigma\tau}}$
for the metric Eq.~(\ref{eq:6a}) is
\begin{eqnarray}
{\cal K}&=&\frac{ 2\ddot{R}'^2}{R'^2}+\frac{4(D-2)\ddot{R}^2}{R^2}
+\frac{4(D-2)\dot{R}^2\dot{R}'^2}{R'^2R^2}\nonumber\\
&+&\frac{2(D-2)(D-3)\dot{R}^4}{R^4}.\label{eq:21a}
\end{eqnarray}
It can be certified that the Kretschmann scalar is finite on the
initial data surface. According to Eq.~(\ref{eq:14a}),
the energy density of  fluid dust sphere is
\begin{eqnarray}
\varepsilon ({t,r}) = \frac{{(D-2)M'}}{{6{R^{D-2}}R'}}.\label{eq:22a}
\end{eqnarray}
Clearly the Kretschmann scalar and the energy density
diverge when $R' = 0$ and $R = 0$. Hence, we have both shell
crossing  and shell focusing singularities.

Shell crossing singularities can be naked, but they are inessential.
Although the mass density and curvature invariants blow up there,
the metric is in fact continuous. This is seen from that the Riemann
curvature tensor and the
energy momentum tensor are well defined, if we replace the
coordinates $(t,r)$ by $(t,R)$. On the other hand, shell focusing
singularities are considered to be the genuine singularities in
space-time manifold \cite{Christodoulou:1984mz}\cite{Deshingkar:1998cb}.
Henceforth, we only concern  the shell focusing singularity here.

In third order Lovelock gravity, equations and solutions of
gravitational collapse are quite different from counterparts in general
relativity. Hence it is necessary  to investigate whether the
evolution of  collapsing dust cloud  leads to the
formation of the shell crossing singularity. We assume each shell satisfies $\dot{R}(0,r)<0$ initially. If $\dot{R}=0$ is satisfied for some
$r$ after the initial moment, this shell ceases to collapse and then bounces ($\dot{R}>0$),
and the shell focusing singularity would not be formed. Indeed, it can be proven that
in the case $\beta\geq0$, $\dot{R}^2-K\geq0$ at the initial moment, the shell focusing singularity
will be inevitably formed.

{\it Proof:} From Eq.~(\ref{eq:15a}), we find that
\begin{eqnarray}
\frac{d(\dot{R}^2)}{dR}=\frac{(7-D)MR^{6-D}-6\alpha\ell^2R-12\ell R^3}{3\beta\ell^2+6\alpha\ell R^2+3R^4},\label{eq:07c}
\end{eqnarray}
where $\ell=\dot{R}^2-K$. Clearly, $\frac{d(\dot{R}^2)}{dR}<0$ if $\beta\geq0$ and $\ell\geq0$. Here we have used the condition $\alpha\geq0$ which is mentioned in the last section.
If $\ell$ is non-negative at the initial surface, it will be always non-negative as increasing $\dot{R}^2$. Hence, $|\dot{R}|$ increase as decreasing $R$ therefore the collapsing of the shell $r=const$ from the finite
initial data $R( {0,r} ) = r$ is accelerated. Thus, $R$ will vanish
in finite proper time for the comoving observer, the shell focusing
singularity is inevitably formed. $\Box$

 Physically, the condition  $l\geq 0$ is satisfied for $\beta\geq\alpha^2$.
For $0<\beta<\alpha^2$, one should take in mind  Eq.(\ref{eq:001c}) with a real solution, which gives a lower limit of $\rho_D$.
 Within this region, the numerical figure show the condition valid, but it is hard to analytically prove it. Moreover, one can show the condition $l\geq0$
 valid forever if it is satisfied at the beginning.
Thus, the final fate of a freely falling
fluid sphere, which have initial density $\varepsilon ( {0,r} )$ and
zero pressure, is a state of infinite energy density and curvature.
%, if $\beta\geq0$ and $\dot{R}(0,r)^2-K\geq0$.
%Furthermore, if $\beta\geq\alpha^2$, it is straightforward to check that $\dot{R}^2-K>0$. Hence, in the case $\beta\geq\alpha^2$, the final fate of the collapse
%is a shell focusing singularity.
In the rest of this paper we will only consider the case that the shell focusing singularity could be formed, and assume $\beta\geq0$.

As demonstrate  above, our solution can
be attached to the outside vacuum solution at the boundary of the dust cloud. The outgoing property
of the central singularity depends on the dominant term of $F(R)$ in vacuum solution for $R \to 0$ \cite{Torii:2005xu}.
If the metric function $F<0$, then the tortoise coordinate defined by
\begin{eqnarray}
R^\star=\int^R F^{-1}dR\nonumber
\end{eqnarray}
has finite negative value, the singularity is space-like. If $F>0$, the singularity is time-like. If $F=0$,
the singularity is null since $|R^\star|\rightarrow\infty$. From Eq.~(\ref{eq:03c}), we find that the metric function
behaves around the central singularity as
\begin{eqnarray}
F(R)\approx1-(\frac{M(r_b)}{\beta R^{D-7}})^{\frac{1}{3}},\nonumber
\end{eqnarray}
where ${r_b}$ is the boundary of the collapsing dust cloud. Since we assume $\beta>0$, if $D>7$, we have $F(R)<0$ hence the singularity is space-like. If $D=7$ and $M(r_b)>\beta$, the singularity is also space-like.
If $D=7$ and $M(r_b)=\beta$, it is null. If $D=7$ and $M(r_b)<\beta$, it is time-like. This conclusion is shown in the Penrose diagram FIG.1. The event horizon of such vacuum solution would be discussed in the next section.

If the energy density is independent on $r$,
namely the energy density and the space are homogeneous, the time of
the formation of the singularity ${t_{SF}}$ is a constant,
as in the general relativity case. A homogeneous
space that is isotropic about some point is maximally symmetric and the
curvature in such a space is a constant. In homogeneous case, the
metric Eq.~(\ref{eq:6a}) takes the form as the Robertson-Walker
metric \cite{Robertson:1935}\cite{Walker:1936}
\begin{eqnarray}
d{s^2} =  - d{t^2} + a( t )^2( {\frac{{d{r^2}}}{{1 + k{r^2}}} +
{r^2}d\Omega _{D-2}^2} ),\label{eq:25a}
\end{eqnarray}
where $k$ is a constant. In order to satisfy Eq.~(\ref{eq:10a}),
the radial coordinate $r$ can be normalized so that $a( 0 ) = 1$.
When the shell hits the shell focusing singularity, the time is
completely determined by $a( t )=0$.
Considering metric Eq.~(\ref{eq:25a}) and $
M( r ) = \frac{6}{(D-1)(D-2)}\varepsilon ( 0 ){r^{D-1}}$, we get
\begin{eqnarray}
\dot{a}(t) =-\sqrt{ k+\frac{a(t)^2}{\beta}[(\frac{\hat{\Theta}+\hat{\Pi}}{2})^{\frac{1}{3}}+(\frac{\hat{\Theta}-\hat{\Pi}}{2})^{\frac{1}{3}}-\alpha]} ,\label{eq:01c}
\end{eqnarray}
where
\begin{eqnarray}
\hat{\Theta}&=&-2\alpha^3+3\alpha\beta +\beta^2\hat{\varepsilon},\nonumber\\
\hat{\Pi}&=&\sqrt{\beta^2(\beta^2 \hat{\varepsilon}^2-4\alpha^3\hat{\varepsilon}+6\alpha\beta \hat{\varepsilon}-3\alpha^2+4\beta) }\nonumber
\end{eqnarray}
with $\hat{\varepsilon}=6\varepsilon(0)a(t)^{1-D}/(D-1)(D-2)$. This equation does not contain the variable $r$, that is, $a( t )$
is a function of $t$ independent on $r$. It implies that every shell
will collapse into the shell focusing singularity at the same time.
Obviously, this conclusion is hold in general relativity \cite{Oppenheimer:1939ue},
Gauss-Bonnet gravity \cite{Jhingan:2010zz} and third order Lovelock case.

From the geometric perspective, it can be proven that in the
homogeneous case the shell crossing singularity is ingoing null
when $D=7$ and is ingoing space-like when $D>7$, as shown in FIG.1.

{\it Proof:} Considering Eq.~(\ref{eq:01c}) we have
\begin{eqnarray}
\frac{6}{(D-1)(D-2)}\varepsilon ( 0 )&=&[\beta{({{\dot{a}^2}-k})^2}
+3\alpha {({{\dot{a}^2} - k} )}a^2 \nonumber\\
&+&3a^4]( {{\dot{a}^2} - k})a^{D-7}.\label{eq:28a}
\end{eqnarray}
Since the energy density $\varepsilon ( 0 )$ have a finite non-negative value,
we find that the factor $a$
behaves as $c(t-t_{SF})^{\frac{6}{D-1}}$ near the
singularity ${t = {t_{SF}}}$, where $c$ is a coefficient. We take the line element
of the FRW solutions with the factor $a
= c(t-t_{SF})^{\frac{6}{D-1}}$ to the conformally flat
form as
\begin{eqnarray}
ds^2 = a^2( t( \bar{t}, \bar{r} ) )b^2( r( \bar{r} )
)(  - d\bar{t}^2 + d\bar{r}^2 + \bar{r}^2d\Omega _{D-2}^2),\label{eq:29a}
\end{eqnarray}
where
\begin{eqnarray}
d\bar{t} &=& \frac{dt}{a(t)b(r)},\quad
\bar{r} = \frac{r}{b(r)}, \nonumber\\
b( r ) &=& \exp (\ln | r | + \ln | \frac{1}{r} + \sqrt
{\frac{1}{{{r^2}}} + k}  |).\nonumber
\end{eqnarray}
The range of $\bar{t} $ for $t \in ( { - \infty ,{t_{SF}}} )$ is
$( { - \infty , + \infty } )$ when $D=7$ and is $(-\infty , \bar{t}_0 )$
for $D>7$, where $\bar{t}_0$ is a
constant. Thus, the shell focusing singularity
is ingoing null for $D=7$, while it is ingoing space-like
for $D>7$. $\Box$

%%%%%%%%%%%%%%%%%%%%%%%%%%%%%%%%%%%%%%%%%%%%%%%%%%%%%%%%%%%%%%%%%%%%%%%%%%%
\begin{figure}
\centering \subfigure[$D\geq8$]{
\label{fig:subfig:A} %% label for second subfigure
\includegraphics{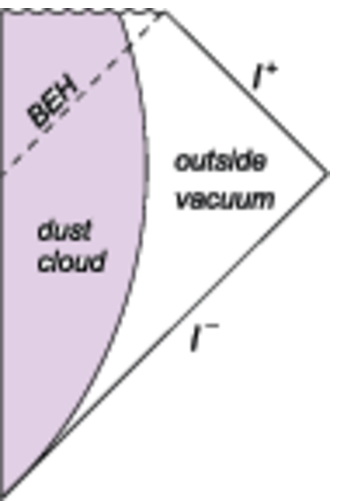}}%
\hfill%
\subfigure[$D=7, M(r_b) > \beta$]{
\label{fig:subfig:B} %% label for second subfigure
\includegraphics{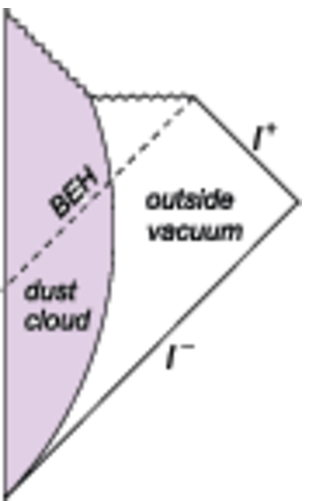}}
\hfill%
\subfigure[$D=7, M(r_b) = \beta$]{
\label{fig:subfig:C} %% label for second subfigure
\includegraphics{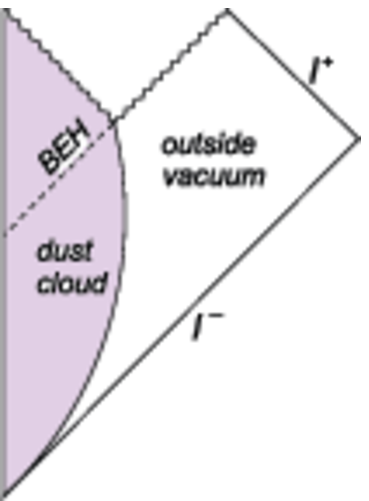}}
\hfill%
\subfigure[$D=7, M(r_b) < \beta$]{
\label{fig:subfig:B} %% label for second subfigure
\includegraphics{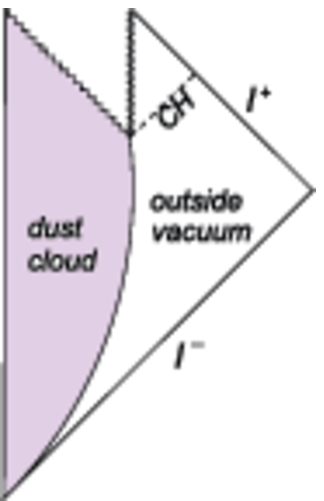}}
\caption{ Penrose diagram of the homogeneous collapse of a
spherically symmetric dust cloud in third order Lovelock gravity.
Zigzag lines represent the shell
focusing singularities, $I^{+(-)}$ corresponds to the future (past)
null infinity, BEH and CH stand for the black hole event horizon and
the Cauchy horizon, respectively. }
\label{fig:subfig} %% label for entire figure
\end{figure}

\section{Horizon and trapped surface\label{44s}}
\indent

An important construction in
general relativity is that of the trapped surface, which is
indispensable in proving null-geodesics incompleteness in context of
gravitational collapse \cite{Hawking:1973uf}. In general relativity,
the trapped surface is defined as a ${C^2}$ closed space-like
two-surface that two families of null geodesics orthogonal to
this surface are converging. Physically, it captures the notion of
trapping by implying that if two-surface $\Gamma(t,r)$ ($t,r=const$) is a
trapped surface, then its entire future development lies behind a
horizon. The apparent horizon is the outermost marginally trapped
surface for the outgoing null geodesics \cite{Hayward:1993wb},
therefore the trapped surface could not be formed during collapse
without the occurrence of the apparent horizon. The main advantage
of working with the apparent horizon is that it is local in time and
can be located at a given space-like hypersurface. Instead, the
event horizon coincide in case of static or stationary space-time,
it is non-local. Moreover, in the vacuum region the apparent horizon coincides with the event horizon of the vacuum solution \cite{Christodoulou:1984mz}. That is, without the presence of
apparent horizons there is no event horizon.

Demanding the presence
of the trapped surface in our spherically symmetric case implies
\begin{eqnarray}
{g^{\mu \nu }}{R_{,\mu }}{R_{,\nu }}=  - {\dot{R}^2} +
\frac{{{R'^2}}}{{{A^2}}} < 0,\label{eq:30a}
\end{eqnarray}
and the condition for the existence of the apparent horizon with
outward normals null is
\begin{eqnarray}
{g^{\mu \nu }}{R_{,\mu }}{R_{,\nu }} =  - {\dot{R}^2} +
\frac{{{R'^2}}}{{{A^2}}} = 0.\label{eq:31a}
\end{eqnarray}
Using Eq.~(\ref{eq:8a}), the apparent
horizon condition becomes
\begin{eqnarray}
 \dot{R}^2-K= 1,\label{eq:08c}
\end{eqnarray}
Combing Eq.~(\ref{eq:15a}) and Eq.~(\ref{eq:08c}), we obtain
\begin{eqnarray}
{R_{AH}}(t_{AH}(r),r)=\sqrt{\frac{-3\alpha + \Psi}{6}},\label{eq:32a}
\end{eqnarray}
where
\begin{eqnarray}
\Psi=\sqrt{12M(r)R_{AH}(t_{AH}(r),r)^{7-D}+9\alpha^2-12\beta}.\label{eq:32a}
\end{eqnarray}
Clearly, coupling constants $\alpha $ and $\beta$ produces a change in the
location of apparent horizons. Such a change could have a signification
effect in the dynamical evolution of these horizons. It has been
shown that in the 5D Gauss-Bonnet gravity case, positive $\alpha$
forbids apparent horizon from reaching the coordinate center thereby
making the singularity massive and eternally visible
\cite{Jhingan:2010zz}, which is forbidden in the corresponding
general relativistic scenario and $D\geq6$ Gauss-Bonnet
gravity \cite{Christodoulou:1984mz}\cite{Maeda:2006pm}\cite{Cooperstock:1996yv}.
In our case, positive $\alpha $ leads to noncentral naked
singularities when $D=7$. From Eq.~(\ref{eq:32a}), we find the condition
that the apparent horizon is earlier than singular shell ($R_{AH}(t_{AH}(r),r)>0$) for
positive $\alpha $ in 7D case is $M( r ) > \beta$.
Oppositely, Eq.~(\ref{eq:32a}) could not be satisfied before the
formation of the singularity in the 7D space-time in the case
$M( r ) < \beta$.
Furthermore, the apparent horizon touches the singularity
when $M( r ) = \beta$. Thus,
there is no shell can reach the apparent horizon if
\begin{eqnarray}
M( {r_b} )<\beta,\label{eq:33a}
\end{eqnarray}
with $r_b$ the boundary of the dust cloud. It implies if the mass
function $M( {r_b} )$ takes sufficiently small value,
the shell focusing singularity is eternally visible from
infinity during collapse, and leave open even the weak form of the
CCH.

The condition of the formation of the eternally visible shell
focusing singularity in 7D case is completely determined by the initial data of
the energy density. With the help of Eq.~(\ref{eq:14a}), we can find
the initial data of the energy density that
condition Eq.~(\ref{eq:33a}) requires. For homogeneous case, such
initial data satisfies
\begin{eqnarray}
\frac{{\varepsilon ( 0 )r_b^6}}{5} < \beta.\label{eq:34a}
\end{eqnarray}
We can consider a more realistic model that
${\varepsilon _0}[ {1 - {{( {\frac{r}{{{r_b}}}} )}^n}}]$,
which is a density profile where energy density decreases as an
observer move away from the center, as is expected inside a star.
This form of initial data leads to
\begin{eqnarray}
\frac{{n{\varepsilon _0}r_b^6}}{{5( {n + 6} )}} < \beta.\label{eq:35a}
\end{eqnarray}

On the other hand, it is clear to see that in the
$D\geq8$ space-time, apparent horizons lies earlier than the
singularity unless $r=0$. Thus, the solution forbids the formation of
the naked singularity except the singular point at the coordinate center,
as same as in the general relativity and the Gauss-Bonnet gravity. As pointed by Christodoulou,
the center singularity can be naked for suitable initial data of $\epsilon(0,r)$ in the inhomogeneous case \cite{Christodoulou:1984mz}.
The physical picture captured here is that the coordinate center hits the singularity
so early that other shells have not reached apparent horizons at this moment, hence the light from the singular coordinate center can escape to the infinity.

In Gauss-Bonnet gravity, the formation of the eternally visible shell
focusing singularity is forbidden in $D\geq6$ space-time, and is permitted when $M<\alpha$ and $\alpha>0$ in the 5D case. Comparing results in second and third order Lovelock gravity, one can conjectures: for $n$ order Lovelock gravity, such a naked singularity
does not exist in $D\geq2n+2$ space-time, and would be formed when $M<\alpha_n$ in $D=2n+1$ case, where $\alpha_n$ is the coefficient of the highest order Lovelock term and is positive defined.

Now we consider the event horizon of the vacuum solution which describes the final fate of the collapse. Based on Eq.~(\ref{eq:03c}), we find the metric function $F(R)$
satisfies
\begin{eqnarray}
M(r_b)&=&R^{D-7}\Big[\beta(1-F(R))^3
+ 3\alpha (1-F(R))^2{R^2}\nonumber\\
&+&3(1-F(R)){R^4} \Big]
.\nonumber
\end{eqnarray}
The event horizon occurs when $F(R_{EH})=0$, thus we obtain
\begin{eqnarray}
M(r_b)=R_{EH}^{D-7}(\beta
+ 3\alpha {R_{EH}^2}
+3{R_{EH}^4} )
.\nonumber
\end{eqnarray}
This equation leads to that
\begin{eqnarray}
{R_{EH}}=\sqrt{\frac{-3\alpha + \sqrt{12M(r_b)R_{EH}^{7-D}+9\alpha^2-12\beta}}{6}}.\nonumber
\end{eqnarray}
Comparing this equation to Eq.~(\ref{eq:32a}), it is obvious to see the condition for the existence of the event horizon is the same to the condition for the presence of the
apparent horizon, which is: it exist in the $D\geq8$ case, and in the 7D case with $M(r_b)>\beta$. If $M(r_b)=\beta$
in 7D space-time, the event horizon touches the singularity at the center of the vacuum solution coordinate. This result is also shown in FIG.1. Thus, the $D=7$ and $M(r_b)<\beta$ case indicates the existence
of naked singularities which are eternally visible from infinity, and violates even the weak form of CCH.
Such naked singularities are forbidden in standard general relativity, and is allowed in collapse in 5D second order Lovelock gravity with positive $\alpha$ \cite{Maeda:2006pm}.
Such conclusion coincides with the result which is obtained by investigating the apparent horizon and the outgoing property of the singularity.

\section{Strength of singularity\label{55s}}
\indent

In the case of formation of a naked singularity in gravitational
collapse, one of the most significant aspects is the strength of
such a singularity in terms of the behavior of the gravitational
tidal forces in its vicinity \cite{Tipler:1977zza}. The importance
of the singularity strength lies in the fact that even if a
naked singularity occurs, if it is gravitationally weak in some
suitable sense, it may not have any physical implications and it may
perhaps be removable by extending the space-time through the same
\cite{Clarke:1994cw}. The gravitational strength of a space-time
singularity is characterized in terms of the behavior of the
linearly independent Jacobi fields along the time-like or null
geodesics which terminate at the singularity. In particular, a
causal geodesic $\gamma ( s )$, incomplete at the affine parameter
value ${s = {s_0}}$, is said to terminate in a strong curvature
singularity at ${s = {s_0}}$, if the volume three-form $V( s ) =
{Z_1}( s ) \bigwedge {Z_2}( s ) \bigwedge {Z_3}( s )$ defined as a
two-form in the case of a null geodesic ) vanishes in the limit as
$( {s \to {s_0}} )$ for all linearly independent vorticity free
Jacobi fields ${Z_1}( s ),{Z_2}( s ),{Z_3}( s )$ along $\gamma ( s
)$. A sufficient condition for a strong curvature singularity in 4D space-time is
that in the limit of approach to the singularity, we must have along
at least one causal geodesic $\gamma ( s )$ \cite{Deshingkar:1998cb}\cite{Clarke:1985},
\begin{eqnarray}
\mathop {\lim }\limits_{s \to {s_0}} {( {s - {s_0}} )^2}\psi =
\mathop {\lim }\limits_{s \to {s_0}} {( {s - {s_0}}
)^2}{R_{ab}}{V^a}{V^b} > 0,\label{eq:37a}
\end{eqnarray}
where ${V^a}$ is the tangent vector to the geodesic. Essentially, the idea
captured here is that in the limit of approach to such a
singularity, the physical objects get crushed to a zero size, and so
the idea of extension of space-time through it would not make sense,
characterizing this to be a genuine space-time singularity. This condition has been applied to higher
 dimensional case by Ghosh, Beesham and Jhingan \cite{Ghosh:2000}\cite{Jhingan:2010zz}. Here we follow them to assume that such condition can be applied to our case.

We consider radial time-like causal geodesics ${U^\mu} =
\frac{{d{x^\mu}}}{{d\tau }}$ with the marginally bound case $K=0$,
here the affine parameter $\tau $ is the proper time along particle
trajectories. According to such definition, ${U^\mu}$ satisfies ${U^\mu}{U_\mu}= -1$, that is,
\begin{eqnarray}
- {( {\frac{{d{x^t}}}{{d\tau }}} )^2} + {R'^2}{(
{\frac{{d{x^r}}}{{d\tau }}} )^2} =  - 1.\label{eq:38a}
\end{eqnarray}
Using the geodesic equation, we have
\begin{eqnarray}
\frac{d^2x^t}{d\tau^2}+
\frac{1}{2}(R'^2)_{,t}(\frac{dx^r}{d\tau})^2=0.\label{eq:39a}
\end{eqnarray}
Substituting Eq.~(\ref{eq:38a}) into Eq.~(\ref{eq:39a}), we obtain
that radial time-like geodesics must satisfy
\begin{eqnarray}
\frac{{d{U^t}}}{{d\tau }} + \frac{{\dot{R}'}}{{R'}}( {{{( {{U^t}}
)}^2} - 1} ) = 0.\label{eq:41a}
\end{eqnarray}
The simplest solution is the worldline of a freely falling particle,
which is ${U^\mu} = \frac{{d{x^\mu}}}{{d\tau }} = \delta _t^a$. In
terms of proper time we can describe it as
\begin{eqnarray}
{t_{SF}}( r ) - t = {\tau _0} - \tau .\label{eq:42a}
\end{eqnarray}
Eq.~(\ref{eq:33a}) shows that the naked shell focusing singularity in 7D space-time
occurs when the region of the collapsing dust is sufficiently small.
Moreover, as mentioned earlier, in $D\geq8$ space-time only the central singularity can be naked.
Hence, our purpose is to discuss the strength of the central singularity.
We consider the expansion of $\varepsilon(r)$ near $r=0$
\begin{eqnarray}
\varepsilon(r) = \sum\limits_{n = 0}^{ + \infty } {{\varepsilon_n}{r^n}}
\simeq {\varepsilon_0},\label{eq:43a}
\end{eqnarray}
it specify that $R(t,r)$ behaves as ${R_0}( t )r$ (homogeneous case) near the coordinate center,
where ${R_0}( t )$ is a function of $t$ and vanishes at $t={t_{SF}}$.
Thus, we get
\begin{eqnarray}
\psi  &=& {R_{ab}}{U^a}{U^b} =  - \frac{{(D-1){\ddot{R}_0}}}{{{R_0}}}.\label{eq:44a}
\end{eqnarray}
Clearly,
\begin{eqnarray}
\mathop {\lim }\limits_{\tau  \to {\tau _0}} \frac{{R_0(t)}}{{\tau -
{\tau _0}}}=\mathop {\lim }\limits_{\tau  \to {\tau_0}}
\frac{{R_0(t)-R_0(t_{SF})}}{{t - {t _{SF}}}}=\mathop {\lim
}\limits_{\tau \to {\tau_0}} {\dot{R}_0(t)},\nonumber
\end{eqnarray}
thus we have
\begin{eqnarray}
&&\mathop {\lim }\limits_{\tau  \to {\tau _0}} ( {\tau  - {\tau _0}}
)^2=\mathop {\lim }\limits_{\tau  \to {\tau _0}} \frac{R_0^2}{\dot{R}_0^2},\nonumber\\
&&\mathop {\lim }\limits_{\tau  \to {\tau _0}}( {\tau  - {\tau _0}}
)^2\psi =\mathop {\lim }\limits_{\tau  \to {\tau _0}} -(D-1)\frac{R_0\ddot{R}_0}{\dot{R}_0^2}. \label{eq:46a}
\end{eqnarray}
From Eq.~(\ref{eq:001c}), we obtain that
\begin{eqnarray}
&&\mathop {\lim }\limits_{\tau  \to {\tau _0}}\dot{R}_0^2=\mathop {\lim }\limits_{\tau  \to {\tau _0}} (\frac{M_0}{\beta})^{\frac{1}{3}}R_0^{\frac{7-D}{3}},\nonumber\\
&&\mathop {\lim }\limits_{\tau  \to {\tau _0}}\ddot{R}_0=\mathop {\lim }\limits_{\tau  \to {\tau _0}} \frac{7-D}{6}(\frac{M_0}{\beta})^{\frac{1}{3}}R_0^{\frac{4-D}{3}}\label{eq:45a}
\end{eqnarray}
where ${{M_0}}$ is defined as ${M_0} = \mathop {\lim }\limits_{r \to
0} \frac{{M( r )}}{{{r^{D-1}}}}$ therefore has finite value.
Substituting Eq.~(\ref{eq:45a}) into Eq.~(\ref{eq:46a}) we have
\begin{eqnarray}
\mathop {\lim }\limits_{\tau  \to {\tau _0}} ( {\tau  - {\tau _0}}
)^2\psi  = \frac{{(D-1)(D-7)}}{6},\label{eq:48a}
\end{eqnarray}
the strong curvature condition Eq.~(\ref{eq:37a}) is not satisfied on
the singularity near $r=0$ in the 7D space-time, while it is satisfied in the $D\geq8$ space-time.

Hence, the central shell focusing singularity is gravitational strong
in the $D\geq8$ space-time and may be gravitational weak in the 7D space-time since we chosen a special solution of Eq.~(\ref{eq:41a}). The
difference between two cases is that the singularity can be
eternally visible in the 7D case while it is forbidden in the $D\geq8$
case. That is, the singularity is gravitational strong when the
solution represents the black hole formation for arbitrary initial
data. If a naked singularity is gravitational weak, it may not have
any significant physical consequences so may not be a serious threat
to the CCH. Thus, the naked singularity in the $D\geq8$ case
violates the CCH seriously, as in the general relativity
\cite{Christodoulou:1984mz}\cite{Deshingkar:1998cb}, while the naked
singularity may not be regarded as an essential counter example to the CCH in
the 7D case.

Nevertheless, such a central singularity in the 7D case is gravitational strong in the
general relativistic and second order Lovelock gravity. In the general relativistic limit $\alpha \to 0, \beta \to 0$, Eq.~(\ref{eq:37a}) takes the form
\begin{eqnarray}
\mathop {\lim }\limits_{\tau  \to {\tau _0}} {( {\tau  - {\tau _0}}
)^2}\psi  = \mathop {\lim }\limits_{\tau  \to {\tau _0}} {( {\tau -
{\tau _0}} )^2}[ {\frac{{(D-1)(D-3){M_0}}}{{6R_0^{D-1}}}} ],\nonumber
\end{eqnarray}
with
\begin{eqnarray}
\mathop {\lim }\limits_{\tau  \to {\tau _0}} {\left( {\tau  - {\tau
_0}} \right)^2} = \mathop {\lim }\limits_{\tau  \to {\tau _0}}
\frac{{3R_0^{D-1}}}{{{M_0}}}.\nonumber
\end{eqnarray}
Thus,
\begin{eqnarray}
\mathop {\lim }\limits_{\tau  \to {\tau _0}} ( {\tau  - {\tau _0}}
)^2\psi  =  \frac{(D-1)(D-3)}{2}  > 0,\label{eq:51a}
\end{eqnarray}
the strong curvature condition is satisfied in $D>3$ space-time.
In second order Lovelock gravity limit $\beta \to 0$, we find that
\begin{eqnarray}
\mathop {\lim }\limits_{\tau  \to {\tau _0}} ( {\tau  - {\tau _0}}
)^2\psi  =  \frac{(D-1)(D-5)}{4}  > 0,\label{eq:51c}
\end{eqnarray}
the strong curvature condition is satisfied in $D>5$ space-time.
These results indicate that the  Lovelock interaction weaken the strength of the singularity. Moreover,  a generic expression for $n$ order Lovelock gravity could be
\begin{eqnarray}
\mathop {\lim }\limits_{\tau  \to {\tau _0}} ( {\tau  - {\tau _0}}
)^2\psi  =  \frac{(D-1)(D-2n-1)}{2n}  > 0.
\end{eqnarray}
It shows that the strong curvature condition is satisfied in $D>2n+1$ space-time.

\section{conclusions and discussions\label{66s}}
\indent

In this paper, we have investigated the gravitational collapse of a
spherically symmetric dust cloud in $D\geq7$ space-time in third order
Lovelock gravity without a cosmological constant. From field
equations given by the third order Lovelock action, we discussed the
solutions of three families $K( r ) > 0$, $K( r ) < 0$ and $K( r ) =
0$. We discussed the final fate of the collapse, and gave a condition for the formation of the shell focusing singularity.
We also analysed the global structure of the space-time and gave the
Penrose diagram for the homogeneous case. It turns out that, the
contribution of high order curvature corrections has a profound
influence on the nature of the singularity, and the whole physical
picture of the gravitational collapse changes drastically. High
order Lovelock terms alters the course of collapse and the time of
formation of singularities, modifies apparent horizon formation and
the location of apparent horizons, and changes the strength of
singularities.

The most attractive consequence is that an massive naked shell
focusing singularity is inevitably formed in 7D space-time, which is quite different from that in the general
relativity and in the Gauss-Bonnet gravity. However, as we shown,
the strength of the
naked singularity in 7D case is weaker than that in the general
relativistic limit,
therefore this may not be a serious threat to the CCH. On the other hand,
unlike the 7D case, there is a
serious threat to the CCH caused by the naked, gravitational strong
singularity in the $D\geq8$ inhomogeneous collapse, thus the CCH which is violated in
general relativity could not be protected in Lovelock gravity.

When  revising our  paper in lines with reviewer's suggestions, we have received a paper by
 Ohashi, shiromizu and Jhingan\cite{Ohashi:2011} discussing  the gravitational collapse in the Lovelock theory with arbitrary
order  which partially covers the  result of present paper. In fact, they did not investigate
the ingoing and outgoing properties of the singularity, the Penrose diagram, and whether the condition of strong singularity is satisfied.
One may study such properties in a general Lovelock gravity similarly.

{\bf Acknowledgment }
This work has been supported by the Natural Science Foundation of China
under grant No.10875060, 10975180 and 11047025.

\end{document}